\documentclass[twocolumn]{IEEEtran}
\usepackage{color}
\usepackage{url}
\usepackage{cite}
\usepackage{graphicx}
\usepackage[breaklinks=true,colorlinks=true,linkcolor=black,urlcolor=black,citecolor=black]{hyperref}

\begin{document}
	\title{\bf From $\mu_0$ to $e$: A Survey of Major Impacts for Electrical Measurements in Recent SI Revision}
	\author{Shisong Li$^\dagger$,~\IEEEmembership{Senior Member,~IEEE,}
		Qing Wang,~\IEEEmembership{Senior Member,~IEEE,}\\
		Wei Zhao, Songling Huang,~\IEEEmembership{Senior Member,~IEEE}
		\thanks{Shisong Li and Qing Wang are with the Department of Engineering, Durham University, Durham DH1 3LE, United Kingdom. Wei Zhao and Songling Huang are with the Department of Electrical Engineering, Tsinghua University, Beijing 100084, P. R. China.} 
		\thanks{$^\dagger$ Email: leeshisong@sina.com.}
		\thanks{Manuscript submitted to {\it IEEE Trans. Instrum. Meas.} (R2).}}
	\maketitle
	
	\begin{abstract}
		A milestone revision {\color{black}of} the International System of Units (SI) was made at the 26th General Conference on Weights and Measures that four of the seven SI base units, i.e. kilogram, ampere, kelvin, and mole, are redefined by fundamental physical constants of nature. The SI base unit founding the electrical measurement activities, i.e. ampere, is defined by fixing the numerical value of the elementary charge to $e=1.602\,176\,634\times10^{-19}$C. For electrical measurement, several major adjustments, mostly positive, are involved in this SI revision. In this paper, the main impacts of the new SI for electrical measurement activities are surveyed under the new framework.
	\end{abstract}
	
	\IEEEpeerreviewmaketitle
	
	\section{Introduction}
\begin{figure}
		\centering
		\includegraphics[width=0.5\textwidth]{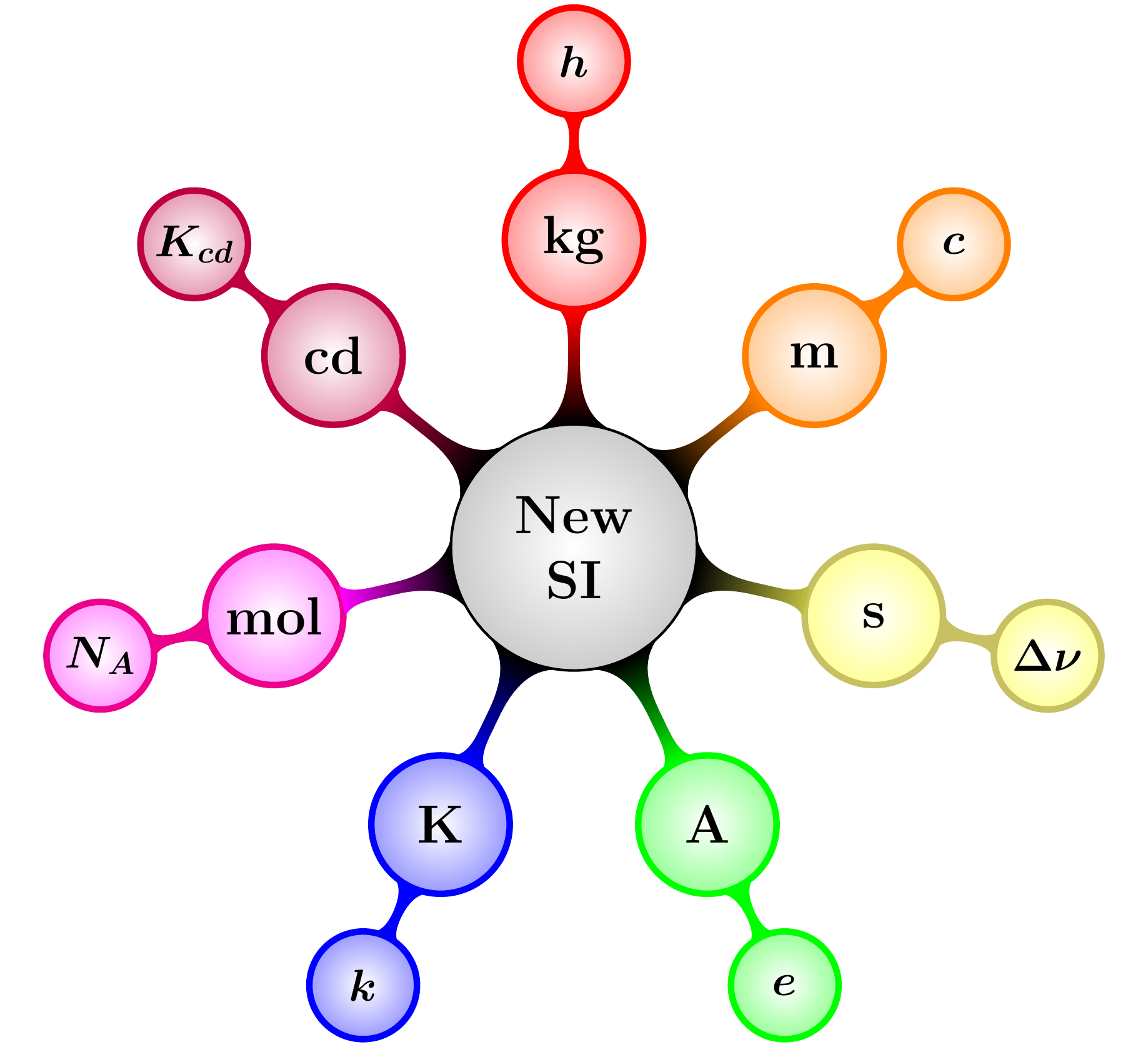}
		\caption{Fundamental physical constants and base units in the revised SI. The outer circles present the seven physical constants for defining the SI base units: The hyperfine transition frequency of Cs atoms, $\Delta\nu=9~192~631~770$\,Hz, the speed of light in vacuum $c=299~792~458$\,m/s, the Planck constant $h=6.626~070~15\times10^{-34}$\,Js, the elementary charge $e=1.602~176~634\times10^{-19}$\,C, the Boltzmann constant $k=1.380~649\times10^{-23}$\,J/K, the Avogadro constant $N_A=6.022~140 76\times1023$\,/mol, and the Luminous efficacy $K_{cd}=683$\,lm/W are used to define second(s), metre(m), kilogram(kg), ampere(A), kelvin(K), mole(mol), and candela(cd). These newly determined numerical values were obtained by the Committee on Data for Science and Technology (CODATA) in 2017 based on a weighted mean of different experimental output worldwide \cite{codata17}.}
		\label{fig01}
	\end{figure}

	\IEEEPARstart{A}{t} the 26th General Conference on Weights and Measures (CGPM, 2018), member states of the Metre Convention came into an agreement that four SI base units, i.e. kilogram, ampere, mole, and kelvin, were to be respectively redefined by fixed numerical values of the Planck constant $h$, the elementary charge $e$, the Avogadro constant $N_A$ and the Boltzmann constant $k$ \cite{cgpm2018}. Note that in this revision, not the four base units are fixed, but the defining physical constants have lost their uncertainties. All units, both base units and derived units, are now derived from these physical constants (as was in fact already the case for the meter, second and candela). This SI revision, which has been applied in practice worldwide since May 20, 2019, is a significant achievement for fundamental metrology. The SI base units, as shown in Fig.\ref{fig01}, for the first time, are defined completely by physical constants of nature, which will found a long-term stable, space and time independent unit system to ensure the traceability of different measurement activities.
	
	\begin{table*}[tp!]
		\centering
		\caption{1990 conventional and newly fixed values for $h$ and $e$.}
		\begin{tabular}{cccc}
			\hline\hline
			&1990 value    &fixed value    & $(X/X_{90}-1)\times10^9$\\
			\hline
			$h/10^{-34}$(Js)    &6.62606885436132    &6.62607015        &195.54\\
			$e/10^{-19}$(C)     &1.60217649161227    &1.602176634    &88.87\\
			\hline\hline
		\end{tabular}
		\label{tab1}
	\end{table*}
	
	\begin{table}[tp!]
		\centering
		\caption{The electrical unit scale change in the new SI compared to the 1990 conventional system. }
		\begin{tabular}{cccc}
			\hline \hline
			Unit & Symbol & $f(h,e)$ & $(X/X_{90}-1)\times10^9$ \\
			\hline
			
			watt    &    W    & $h$&    195.54    \\
			joule    &    J    & $h$&    195.54    \\
			ohm     &    $\Omega$    & $h/e^2$&    17.79    \\
			siemens    &    S    &$e^2/h$&    -17.79    \\
			farad    &    F    &$e^2/h$&    -17.79    \\
			henry    &    H    &$h/e^2$&    17.79    \\
			ampere    &    A    &$e$&    88.87    \\
			coulomb    &    C    &$e$&    88.87    \\
			volt    &    V    &$h/e$ &    106.67    \\
			weber   &   Wb  &$h/e$  &   106.67\\
			tesla    &    T    &$h/e$ &    106.67\\
			\hline \hline
		\end{tabular}
		\label{tab2}
	\end{table}
	
	For seven decades since 1948, the definition of the SI base unit for electrical current, i.e. ampere, had been linked to the electromagnetic force produced by an ideal geometry (two straight parallel conductors of infinite length, of negligible circular cross-section, and placed 1\,m apart in vacuum). Without the background knowledge in electrical measurement area, the above ideal conditions caused difficulty and confusion in understanding such a definition. Although in practice the current can be precisely deduced by Ohm's law based on quantum voltage and resistance standards \cite{jvs,qhr}, the value obtained was under an independent unit system used only in the electrical measurement community, i.e. the 1990 conventional electrical system \cite{1990,1988}, whose unit size differs from the SI value by up to a few parts in $10^{7}$. The reason for the 1990 system splitting off from the SI is that the quantum standards can realize {\color{black}the} volt and {\color{black}the} ohm with a reproducibility better than $10^{-9}$, while their uncertainties were dominated by the uncertainty of related fundamental constants, i.e. $e$ and $h$, at only $10^{-7}$. In the revised SI,  the ampere is redefined by the elementary charge $e$ fixed at a SI defined value {\color{black}of} $1.602~176~634\times10^{-19}$\,C \cite{codata17}. This new definition eliminates the above-listed drawbacks from the root and brings electrical measurements back to the SI system. In addition, the SI revision introduces new electrical applications in other areas. For example, the mass and force at different scales can be realized or calibrated via electrical measurements \cite{kb16,smallforce}. As another example, electrical measurements can also help for realizing the base unit kelvin and measuring temperatures in the new SI \cite{te,jnt}.

	Although wide topics on the 2019 SI revision have been discussed elsewhere, e.g. \cite{review1,review2,pt,review3,ss19,ss20}, here we aim to narrow the discussion and specialize it in the electrical measurement field. {\color{black}Furthermore}, it is expected to summarize materials of existing discussions, e.g. \cite{review4,cce}, with the latest update, and as a result, to deliver a concise overview of the major changes in electrical measurements related to the 2019 SI revision to general audiences in the electricity and magnetism community, especially for {\color{black}non-metrologists}. The rest of the paper is organized as follows: In section \ref{sec02}, the measurement data contributing to the ampere redefinition are reviewed and the electrical unit scale change is hence deduced. In section \ref{sec03}, the change of two physical constants widely appearing in electrical laws and measurements, i.e. the vacuum permeability $\mu_0$ and the vacuum permittivity $\varepsilon_0$, are discussed. In section \ref{sec03}, we first recall the old electrical metrology and traceability system and then discuss the change for electrical unit realizations related to the SI revision. In section \ref{sec05} and section \ref{sec06}, we respectively summarize the future mass, force metrology and the related temperature realizations, as applications of electrical measurements.

	\section{Electrical unit scale change}
	\label{sec02}
	
	The first considerable result of the 2019 SI revision is the scale adjustment for electrical units.     
	Since 1990, the conventional electrical system has been introduced to electrical measurement \cite{1990}. The 1990 electrical system employed two conventional values for the Josephson constant ($K_J=2e/h$) and the von Klitzing constant ($R_K=h/e^2$), i.e.
	\begin{eqnarray}
	K_{J-90} &=& \frac{2e_{90}}{h_{90}}=483597.9~\mbox{GHz/V}\nonumber\\
	R_{K-90} &=& \frac{h_{90}}{e^2_{90}}=25812.807~\Omega.
	\label{e1}
	\end{eqnarray}
	As presented in (\ref{e1}), the defined $K_{J-90}$ and $R_{K-90}$ can infer fixed values for the Planck constant and the elementary charge in the 1990 conventional system, i.e. $h_{90}$ and $e_{90}$.  Their values (rounded to 15 significant digits) are compared in Tab. \ref{tab1} with the newly fixed $h$ and $e$ values in the revision. The detailed input data and weighted mean calculation can be found in \cite{codata17}. It can be seen that the newly fixed values of $h$ and $e$, compared to $h_{90}$ and $e_{90}$, increase by 195.54\,ppb (parts per billion or parts in $10^9$) and 88.87\,ppb, respectively. Since each electrical unit can be expressed by a combination of the units of $e$ and $h$, its scale change in the new SI system can be calculated from the $h$, $e$ value adjustments. Tab. \ref{tab2} summarizes the unit scale change of some major electrical units, where $f(h,e)$ denotes the minimum function to realize such a unit by the combination of $h$ and $e$. Note that the last column in Tab. \ref{tab2} presents the unit scale change compared to its 1990 conventional value. A plus sign denotes the unit scale in the 1990 system is smaller than the value in the newly revised SI system. As seen from Tab. \ref{tab2}, the absolute scale change ranges from 17.79\,ppb to 195.54\,ppb for different electrical units. In practice, these scale adjustments are not noticeable in general electrical measurement activities, but for occasions requiring a precision calibration or high-accuracy measurements, these new adjustments can be significant. For example, the Josephson voltage standard (JVS) can produce a 10\,V voltage within an uncertainty below $1\times10^{-9}$ \cite{jvs}, and the international comparison of the quantum Hall resistance (QHR) standards is within a few parts in $10^9$ \cite{qhr}. Therefore, the scaling factors need to be updated and corrected in electrical calibration and measurement systems \cite{cce}.

	\section{Uncertainty of $\mu_0$ and $\varepsilon_0$}
	\label{sec03}
	
	Fundamental constants, defined by physical laws, are subject to experimental measurements, and in many cases, their measurement accuracy is correlated. As a result of physical constant adjustments, the measurement uncertainties among different constants must fit into the new framework. In electricity and magnetism, two constants, the vacuum permeability $\mu_0$ and the vacuum permittivity $\varepsilon_0$, which were both defined in the past as exact numbers, i.e. $\mu_0=4\pi\times10^{-7}$\,H/m, $\varepsilon_0=1/(\mu_0c^2)$ (the speed of light in vacuum $c$ is fixed by the meter definition), will be assigned with uncertainties in the revised SI. This can be seen by writing $\mu_0$ and $\varepsilon_0$ in terms of $h$ and $e$, i.e.
	\begin{equation}
	\mu_0=\frac{2h\alpha}{e^2c},
	\label{mu}
	\end{equation}
	\begin{equation}
	\varepsilon_0=\frac{e^2}{2hc\alpha},
	\label{var}
	\end{equation}
	where $\alpha$ is the fine structure constant. With $h$, $e$, $c$ being fixed in (\ref{mu}) and (\ref{var}), the newly assigned uncertainty for $\mu_0$ and $\varepsilon_0$ is equal to the uncertainty of $\alpha$. The Committee on Data for Science and Technology (CODATA) updates the values of physical constants every four years. According to the latest $\alpha$ adjustment in 2018, the measurement uncertainty of $\alpha$ is about $1.5\times10^{-10}$ \cite{codata18}. Although this uncertainty component is small, the coefficient of some electrical laws, e.g. Maxwell's equations, is no longer exact as in the past and will contain measurement uncertainties. It is pointed out by R. Davis in \cite{cgs} that the 2019 revision will also change the exactness of conversions between SI and CGS systems, because the conversion factor may contain $\mu_0$ or $\varepsilon_0$. Since the Gaussian unit (one of the unit systems in CGS) is still used in electromagnetism, especially in magnetism \cite{mag}, conversion factors of different physical quantities between SI and Gaussian systems, as well as their up-to-date conversion uncertainties, are summarized in Tab. \ref{tab3}. One interesting point is that the elementary charge $e$ is exact in the SI (without uncertainty), but it has an uncertainty of $8\times10^{-11}$ in the Gaussian system.
	A full summary of the uncertainty change of physical constants related to the SI revision is detailed in the Appendix (by updating \cite{pt} with the latest adjustment of physical constants \cite{codata17,codata18}).

	\begin{table*}[tp!]
		\centering
		\caption{The conversion factor and its uncertainty for different electromagnetic units between SI and Gaussian systems. Note that $c$ in the second last column denotes only the numerical value of the speed of light in vacuum, i.e. $c=299~792~458$.}
		\begin{tabular}{cccccc}
			\hline \hline
			Quantity                    &    Symbol        &    SI unit    &    Gaussian unit    &    Conversion factor    &    Uncertainty    \\
			&$X$                            &    $U_\textsc{SI}$    &    $U_\textsc{G}$     &$\frac{X}{U_\textsc{G}}:\frac{X}{U_\textsc{SI}}$&  $u_r\times10^9$\\
			\hline
			electric charge                &    $Q$            &    C            &    {\color{black}statC}            &    $10c$        &    0.08    \\
			electric current            &    $I$            &    A            &     {\color{black}statC}/s        &    $10c$        &    0.08    \\
			electric voltage            &    $U$            &    V            &    statV        &    $10^6/c$    &    0.08    \\
			electric field                &    $E$            &    V/m            &    statV/cm    &    $10^4/c$    &    0.08    \\
			electric displacement field    &    $D$            &    C/m$^2$        &     {\color{black}statC}/cm$^2$    &    $4\pi c/10^3$    &    0.08    \\
			magnetic $B$ field            &    $B$            &    T            &    G            &    $10^4$        &    0.08    \\
			magnetic $H$ field            &    $H$            &    A/m            &    Oe            &    $4\pi/10^3$    &    0.08    \\
			magnetic flux                &    $\phi$    &    Wb            &G$\cdot$cm$^2$    &    $10^8$        &    0.08    \\
			resistance                    &    $R$            &    $\Omega$    &    s/cm        &    $10^5/c^2$    &    0.15    \\
			capacitance                    &    $C$            &    F            &    cm            &    $c^2/10^5$    &    0.15    \\
			inductance                    &    $L$            &    H            &    s$^2$/cm    &    $10^5/c^2$    &    0.15    \\
			\hline \hline
		\end{tabular}
		\label{tab3}
	\end{table*}

	\section{Realization of electrical units}
	\label{sec04}
	
	\begin{figure}
		\centering
		\includegraphics[width=0.5\textwidth]{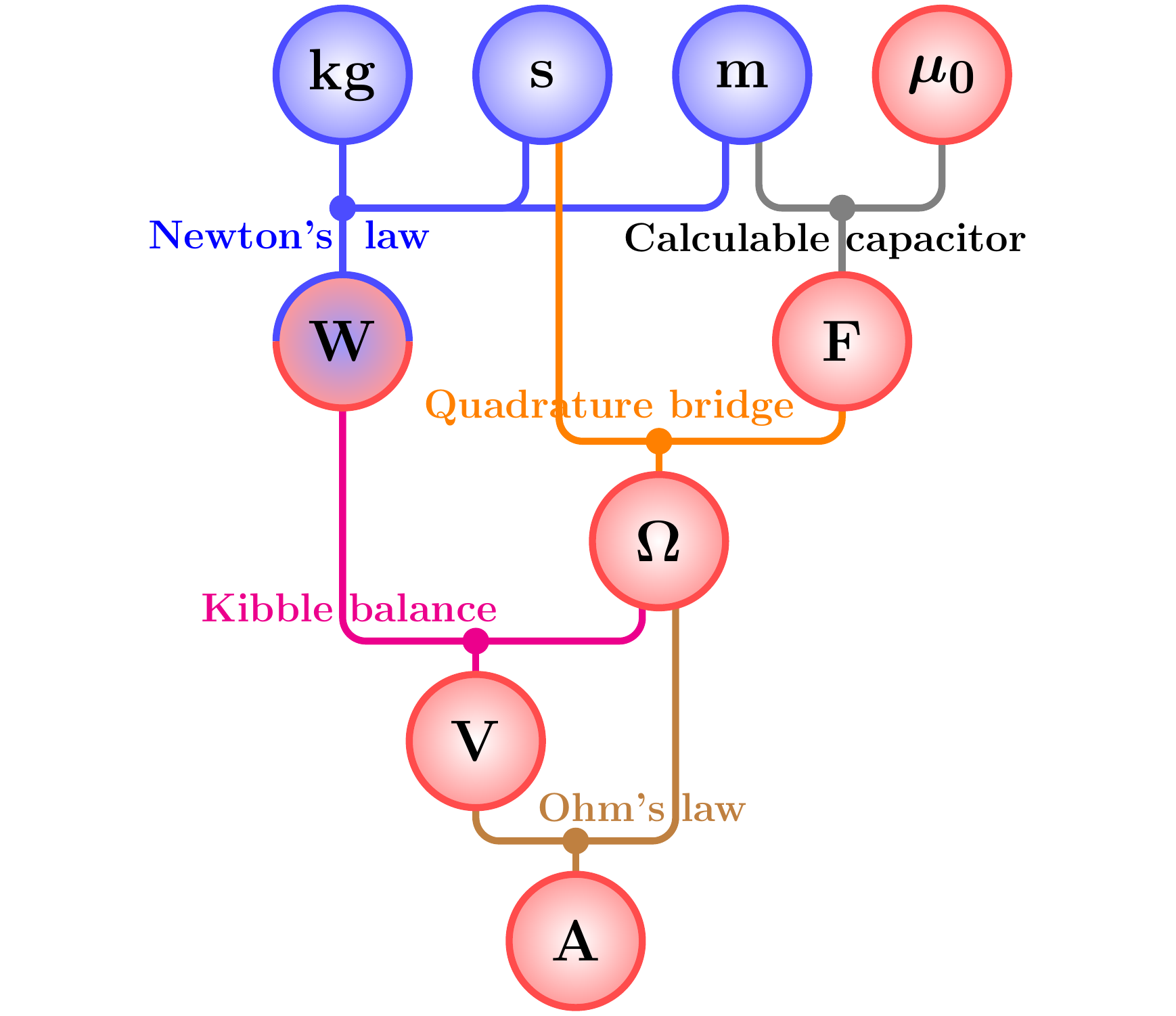}
		\caption{Tractability chain for electrical units in the previous SI system. The red and black stand respectively for electrical unit and mechanical unit. The watt, W, links electrical power to mechanical power by a Kibble balance \cite{kb16}, and hence is marked with both colors.}
		\label{fig02}
	\end{figure}
	
	\subsection{Previous traceability system}
	
	%    \begin{figure}[tp!]
	%        \centering
	%        \includegraphics[width=0.5\textwidth]{cc.pdf}
	%        \caption{}
	%        \label{fig0x}
	%    \end{figure}
	
	To lay a foundation for the following discussions, an overview of the traceability chain for electrical units before the 2019 SI revision is first presented here. The most precise traceability chain for electrical units employed before May 20, 2019, is shown in Fig. \ref{fig02}. The farad, F, was realized by a precision electromechanical instrument, i.e. calculable capacitor, based on an electrostatic theorem discovered by A. M. Thompson and D. G. Lampard at the National Measurement Institute, Australia (NMIA) \cite{cc}. The theorem states that in four infinitely long, parallel conductors in vacuum, the cross-capacitance per unit length between two opposite segments, $c_1$ and $c_2$, meets the following equation
	\begin{equation}
	\exp \left(-\frac{\pi c_1}{\varepsilon_0}\right)+\exp \left(-\frac{\pi c_2}{\varepsilon_0}\right)=1.
	\label{eq3.1}
	\end{equation}
	With symmetrical electrodes, the unit length cross-capacitance $c_1=c_2=c_0$ can be solved by (\ref{eq3.1}), and hence the capacitance is proportional to the effective electrode length $\Delta z$, i.e.
	\begin{equation}
	C=c_0\Delta z=\frac{\ln{2}\Delta z}{\mu_0c^2\pi}.
	\label{eq2.3}
	\end{equation}
	%where $c$ is the speed of light in vacuum.
	Using the Thompson-Lampard theorem, a calculable capacitor links the farad realization to the one-dimensional length measurement, $\Delta z$. This simplification allows a high accuracy capacitance calibration at the pF level. With the great care of machining and aligning the cylinder electrodes, the calculable capacitor can achieve an uncertainty of a few parts in $10^{8}$, e.g. \cite{ccNIM, ccNIST, ccNML,ccLNE}.
	
	The SI value of the ohm was achieved by comparing the resistance value to the capacitance value through the quadrature bridge\cite{qbridge,qbridge2}. The quadrature bridge employs two impedance branches ($R_1, R_2$ and $C_1, C_2$) and measures $R_1R_2$ in terms of $C_1C_2$ as
	\begin{equation}
	\omega^2 C_1C_2R_1R_2=1,
	\label{qb}
	\end{equation}
	where $\omega$ is the measurement angular frequency, $C_1$ and $C_2$ are capacitors calibrated against the calculable capacitor. Note that in order to maintain a low measurement uncertainty and supply an easy link to quantum Hall resistance, the values of $C_1$ and $C_2$ are at nF level, and hence capacitance bridges are required to trace $C_1$ and $C_2$ values to the calculable capacitor \cite{ccLNE, qhr-cc}. With knowing $C_1$ and $C_2$, two further steps are needed to determine $R_1$ and $R_2$: 1) using a resistance bridge to measure the ratio of $R_1$ and $R_2$, and 2) determining the AC/DC difference of $R_1$ and $R_2$ by comparing it against a calculable resistance reference \cite{Haddad}. By minimizing error sources in the above measurement steps, an accuracy of the SI-defined resistance value with a few parts in $10^8$ is achievable.
	
	The realization of the SI volt and ampere was based on the Kibble balance experiment (formerly known as watt balance) \cite{kb16}.  The Kibble balance contains two measurement phases: 1) In the weighing phase, a current-carrying coil is set in a magnetic field and the electromagnetic force of the coil is balanced by the gravitational force of a mass standard, i.e. 
	\begin{equation}
	mg=BLI,
	\end{equation}
	where $B$ is the magnetic field at the coil position, $L$ is the length of the coil wire, $I$ is the current in the coil, $m$ and $g$ represent the mass standard and local gravity acceleration. 2) In the velocity phase, the current is removed and the coil is moved vertically with a velocity $v$, which gives the induced voltage 
	\begin{equation}
	\mathcal{E}=BLv.
	\end{equation}
	Combining two measurement phases, $BL$ can be eliminated and the virtual power balance, 
	\begin{equation}
	mgv=\mathcal{E}I=\mathcal{E}\frac{U}{R},
	\label{kb}
	\end{equation}
	is obtained. Note in the weighing phase, the current $I$ is measured by the voltage drop $U$ on a resistor standard $R$, i.e. $I=U/R$.  
	There are different ways to understand how the volt and the ampere are deduced by (\ref{kb}) and the $R$ measurement. The first approach is to precisely determine the ratio $\gamma=U/\mathcal{E}$ by comparing both $U$ and $\mathcal{E}$ to a JVS reference. Because $\gamma$ has the same value in both 1990 conventional and SI systems, substituting $\mathcal{E}=U/\gamma$ back into (\ref{kb}), the volt and the ampere were realized respectively as
	\begin{equation}
	U=\sqrt{mgv\gamma R},
	\label{ekb}
	\end{equation}
	\begin{equation}
	I=\sqrt{\frac{mgv\gamma}{R}}.
	\label{ekb2}
	\end{equation}
	
	The second understanding is from the view of the physical constant measurement: The Kibble balance was known as an experiment for determining the Planck constant $h$, and the von Klitzing constant $R_K=h/e^2$ can be very precisely deduced from the measurement result of the fine structure constant \cite{codata14}, i.e.
	\begin{equation}
	R_K=\frac{h}{e^2}=\frac{\mu_0c}{2\alpha},
	\end{equation}
	with an uncertainty lower than $1\times10^{-9}$ \cite{alpha,alpha2}. Therefore, $e$ can be determined by $h$ and $\alpha$ (or $h/e^2$) measurements. The Kibble balance idea was to derive electrical power (in terms of $h$, $e$) by treating mechanical power as the reference, and then to deduce volt, ohm, and ampere using quantum standards ($K_J$, $R_K$) and Ohm’s law. We note the above two understandings, respectively from macroscopic measurements and physical constant determinations, are technically equal. The measurement accuracy of a Kibble balance is about a few parts in $10^8$, e.g. \cite{NISTkb, NRCkb, LNEkb}.

	\begin{figure*}
		\centering
		\includegraphics[width=\textwidth]{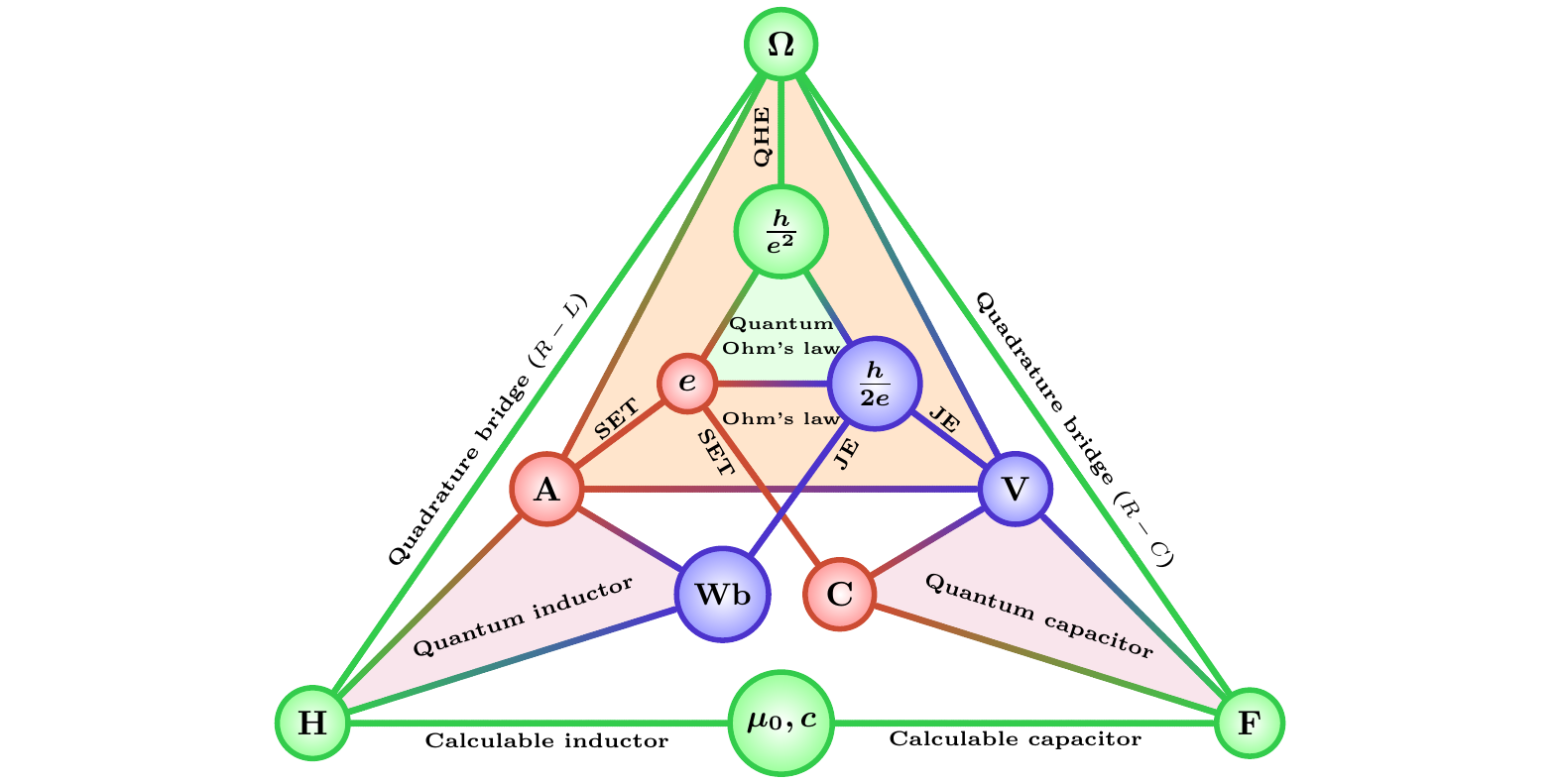}
		\caption{Electrical metrology triangles under the revised SI. Three colors, i.e. green, red and black, respectively denote quantities related to $h/e^2$, $e$, and $h/(2e)$. The inner green triangle presents quantum Ohm's law, while the outer orange is the ordinary Ohm's law. The two pink triangles are respectively the quantum capacitor and the quantum inductor. The outer green triangle shows relations of different kinds of impedance: The $\Omega$ is linked to H and F through quadrature bridges, and the non-resistive quantities, i.e. H and F, can be obtained by calculable standards. }
		\label{fig03}
	\end{figure*}
	
	\subsection{Quantum realization of electrical units}
	The most important impact of the SI revision for electrical measurements is that the SI value for electrical units can be realized in a much easier and straightforward way. Note that quantum standards themselves are not new, but the voltage and the resistance they produced were defined by the 1990 system. The 2019 SI revision sets proper values of $e$ and $h$ without uncertainty to replace the conventional Josephson and von Klitzing constants $K_{J-90}$ ad $R_{K-90}$, and hence brings quantum electrical realizations back into the SI. In the new system, the voltage $U$ can be measured against a Josephson voltage standard (JVS)\cite{jvs}, i.e.
	\begin{equation}
	U=\frac{h}{2e}nf,
	\label{volt}
	\end{equation}
	where $n$ is the number of Josephson junctions used in the measurement and $f$ is the microwave frequency applied to the system. Modern JVS systems employ a programmable bias current so that the sub-arrays of the chip can be chosen, which yields a flexible junction array number $n$ \cite{jvsarray}. As a result, the voltage output (up to 10\,V) can be precisely controlled by fine adjusting $n$ and $f$. Recent comparisons of different JVS systems showed an agreement at 10\,V within $1\times10^{-10}$ \cite{jvsbipm,jvsc2,jvsc3}. Except for DC voltage calibration, the programmable Josephson voltage standard (PJVS) can also be used for AC voltage measurement through a step-wise approximation, e.g. \cite{acjvs}. Meanwhile, a pulse-driven technique, so-called Josephson arbitrary waveform synthesizer (JAWS), has been developed, which offers a different route for realizing AC quantum voltage standards \cite{acpjvs,acpjvs2}. The major difference between two techniques is the RMS amplitude and bandwidth realizations, which {\color{black}in the present art of traceable precision measurements} are: PJVS ($<$7\,V, $<$1\,kHz) and JAWS ($<$2\,V, $<${\color{black}100\,kHz}). {\color{black}The uncertainty of AC voltage comparisons between PJVS and JAWS systems or two JAWS systems is below $1\times10^{-7}$ in the kHz range \cite{acjvscc,nistacpjvsc,acjvsc}.}%The AC voltage comparison of two techniques agrees within a few parts in $10^7$ \cite{}. Direct JAWS comparisons achieved an agreement at the $10^{-8}$ level in a frequency range from 30\,Hz to 2\,kHz\cite{acjvsc}.
	
	In the new SI system, the ohm can be directly realized by quantum Hall resistance (QHR). The resistance $R$ is determined in terms of
	\begin{equation}
	R=\frac{h}{ie^2},
	\label{resistance}
	\end{equation}
	where $i$ is an integer number related to the quantization. The typical comparison uncertainty of different QHR systems is a few parts in $10^9$, e.g. \cite{qhr-comp,qhrc2,qhrc3}. The QHR system can either work conventionally with a single chip at a fixed $i$ value \cite{qhr} or be operated in multiple connecting arrays to realize a fixed value of quantum resistance \cite{qhrarray,qhrarray2}. Recent advances in graphene material research make a QHR standard possible to be operated in a lower magnetic field, higher operating temperature, and higher current density \cite{gqhr,gqhr2,gqhr3}. 
	
	With quantum voltage and resistance standards, a quantum current and hence the ampere can be obtained by combining (\ref{volt}) and (\ref{resistance}), i.e.
	\begin{equation}
	I=\frac{U}{R}=\frac{nife}{2}.
	\label{qampere}
	\end{equation}
	Great progress of such a quantum ampere realization has been made. For example, in \cite{qmulti}, a programmable quantum current source was achieved by integrating the JVS and QHR in the same system, which significantly improves the current measurement accuracy and traceability at the mA range and below.

	In (\ref{qampere}), $n$ and $i$ are known integers, and if $nif/2$ is considered as the counting frequency, then the ampere is realized by counting the elementary charge $e$ following the definition of physical quantity 'current', i.e., the charge going past a given point in one second. Compared to the former definition, the new revision yields a significant simplification for understanding the ampere definition and its realization: One ampere equals a number of $1\mbox{[C]}/e\approx6~241~509~074~460~762~608$ elementary charges going through a given point in one second.
	
	In fact, controlling the elementary charge $e$ was already demonstrated experimentally in the 1980s \cite{set0}. Since then, counting the number of elementary charges in a measurable and repeatable way has been put into experiments for metrology purposes \cite{set, set2}. The counting device, so-called single-electron transistor (SET), allows a single elementary charge $e$, or multiple charges, $Ne$, to go through a potential gate and trigger a counting signal in a fast and controlled speed. By measuring the frequency of the transfer of a fixed number of elementary charges per event, $f$, the ampere can be directly realized by counting $e$ as
	\begin{equation}
	I=Nef.
	\end{equation}
	The principle of the SET realization of the ampere is simple, however, to experimentally realize the ampere with high accuracy is very hard. It is limited by the fact that 1) possible miscounting during the measurement will lead to a non-integer of $N$; 2) the counting frequency is limited below 1\,GHz, and therefore it is difficult to produce a current higher than 100\,pA. At present, the national metrology institute of Germany, Physikalisch-Technische Bundesanstalt (PTB), holds the world accuracy record of ampere realization by SETs. The measurement uncertainty achieved is $1.6\times10^{-7}$ with about 100\,pA, which resulted from a significant measurement robustness improvement using double ultrastable low-noise current amplifiers \cite{ptb-set}.  
	
	As shown in Fig. \ref{fig03}, the three quantum effects: the Josephson effect (JE), the quantum Hall effect (QHE), and the single electron transistor (SET), give a closed triangle of the volt, ohm and ampere realizations. It is known as the 'quantum triangle' \cite{qt}, which allows a quantum check of Ohm's law.  As mentioned above, the quantum triangle experiment is currently limited by the SET leg at a few parts in $10^7$.  
	Fig. \ref{fig03} also extends the quantum measurement triangles to other electrical units.  For example, the SET device allows a determination of coulomb, C, which combines the quantum voltage standard, and yields a quantum farad, F \cite{qc,qc2}.  Similarly, the Josephson effect generates quantum flux and quantum weber, Wb, and with the ampere, it gives a quantum henry, H \cite{qi}. 
	
	It is also noted in Fig. \ref{fig03} that in the revised SI system, the capacitance can be calibrated by multiple paths: 1) Conventional calculable capacitor, 2) DC quantum Hall resistance, then AC/DC difference measurement, and finally $R-C$ quadrature bridge, and 3) quantum capacitor.  A significant advantage, compared to the original calibration through the calculable capacitor and capacitance bridge, is that the new paths 2) and 3) allow a wider and more flexible calibration range for the capacitor, thereby, shorten the length of the traceability chain. For example, path 2) is typically designed for nF capacitance calibration and $\mu$F can be reached if a 100\,$\Omega$ standard resistor is employed. This may lead to significant applications in power and energy measurement areas. 
	It is worth mentioning that researchers have been making efforts to develop an AC quantum Hall resistance standard \cite{acqhr, acqhr2}, which can further reduce the traceability chain of path 2) to AC quantum Hall resistance plus $R-C$ quadrature bridge. 
	
	Traditional inductance standards, realized either by the calculable method \cite{cpl} or the $R-L$ quadrature bridge \cite{rl,rl2}, have a typical measurement uncertainty of a few parts in $10^6$ \cite{inductance}.  The SI revision does not have too much effect on the inductance calibration and henry realization. As mentioned already, the quantum realization of small inductors (at the pH level) may have an application in some precision circuits \cite{qi}.

	\section{Mass and force calibration}
	\label{sec05}
	In the new SI, the Planck constant $h$ has been adopted to redefine the unit of mass, the kilogram, to eliminate the time-dependent drift of the International Prototype of Kilogram (IPK) {\color{black}and provide} long-term stability. Experimental bridges that can link electromagnetic and mechanical power (force), such as the Kibble balance \cite{kb16}, have become the major instruments for mass and force calibration in the revised SI. They extend the application of quantum electrical standards.
	
	As shown in section \ref{sec04}, the Kibble balance was employed for determining $h$, $e$, and hence electrical units by calibrating the electrical power from the mechanical power. But in the new SI, the relation will be upside down, and the electrical quantities will be used to calibrate the mass or force. The Kibble balance becomes one of the two most feasible methods for mass realization at the kilogram level \cite{kb16,silicon} (The other approach is the Avogadro route by counting $^{28}$Si atoms in a highly {\color{black}purified} silicon sphere, details can be found in \cite{silicon2,silicon3}). From (\ref{kb}), the mass $m$ is measured in forms of $\mathcal{E}$, $I$, $g$ and $v$, i.e.
	\begin{equation}
	m=\frac{\mathcal{E}I}{gv}=\frac{\mathcal{E}U}{gvR}. 
	\label{m}
	\end{equation}
	$g$, $v$ are measured by interferometer systems while $U$, $\mathcal{E}$, $R$ are using quantum standards. All quantities can achieve an uncertainty of a few parts in $10^9$, which ensures the mass measurement accuracy within a few parts in $10^8$. It can be seen that the mass realization in (\ref{m}) is a quasi-quantum measurement, and the Kibble balance provides a bridge between the mass and quantum electrical standards.
	
	The Kibble balance also shows great potential for small mass measurement, down to gram level \cite{tbkb}. The Kibble balance for mass calibration under 100\,g can be realized in a simple and tabletop design \cite{tbkb2}. Compared to the conventional calibration path (1kg--100g--10g--1g), a small-mass Kibble balance shortens the measurement chain and has a potential in the future to reduce the uncertainty loss during the transfer. Also, in the new SI, the mass standard value is more flexible and will no longer have to be limited to exact class values. An interesting idea was mentioned in \cite{tbkb3} that both the quantum Hall resistance and the Josephson voltage standard can be integrated into the Kibble balance experiment. One possibility is to use the graphene quantum Hall resistance standard, which can supply the necessary current up to several hundred $\mu$A during the weighing measurement.  
	
	Masses at the milligram level or below are calibrated by electrostatic balances \cite{sf}. A typical electrostatic balance uses a charged capacitor system to produce an electrostatic force that can be precisely measured, i.e.
	\begin{equation}
	F=mg=\frac{U^2}{2}\frac{\partial C}{\partial z},
	\end{equation}
	where $U$ is the voltage applied on the capacitor and $\partial C/\partial z$ the capacitance gradient along the vertical direction $z$. Electrical standards offer an accurate calibration of $C$ (at different $z$ locations) and $U$, and hence the force or mass can be determined. An electrostatic balance can achieve a force calibration uncertainty to a few tens of nN in the range of several hundred $\mu$N\cite{sf2,smallforce}.

	\section{Temperature Measurement}
	\label{sec06}
	Another important application for electrical measurements in the revised SI is to measure the temperature. The new SI adopted a fixed value for the Boltzmann constant $k$ to define the base unit kelvin. The 2017 CODATA final adjustment of $k$ included two experimental results related to electrical measurements \cite{codata17}, i.e. the dielectric-constant gas thermometer (DCGT) \cite{te} and the Johnson noise thermometer (JNT) \cite{jnt}. After $k$ is fixed, DCGT and JNT will serve as primary ways for kelvin realization and temperature measurement. 
	
	The DCGT is based on measuring the pressure-dielectric constant dependence, $p(\varepsilon)$, of an ideal gas (e.g., $^4$He) in a container \cite{kel}. The first-order dependence of $p(\varepsilon)$ allows us to write the temperature $T$ to be measured as 
	\begin{equation}
	T=\frac{\alpha_0p(\varepsilon_r+2)}{3k\varepsilon_0(\varepsilon_r-1)},
	\end{equation}
	where $\varepsilon_r=\varepsilon/\varepsilon_0$ is the relative dielectric constant, and $\alpha_0$ is the static electric dipole polarizability of the gas ({\color{black}e.g., the relative uncertainty} $u_r\approx1\times10^{-7}$ for $^4$He \cite{alpha0}). The key electrical implementation in DCGT is that the dielectric constant $\varepsilon_r$ is calibrated by capacitance measurements, respectively under pressure $p$ and in vacuum, i.e.
	\begin{equation}
	\frac{C(p)-C(0)}{C(0)}=\varepsilon_r-1,
	\label{c}
	\end{equation}
	where $C(p)$ is the capacitance of the capacitor filled with the measuring gas at pressure $p$, and $C(0)$ the capacitance in vacuum. The capacitance, at the pF level, can be well determined at the $10^{-8}$ level by linking the measurement to a reference capacitor via capacitance bridges. Note that for high-precision $\varepsilon$ measurements, a high pressure is required. For example, towards a $10^{-6}$ accuracy, $p$ needs to be about 70 times of the atmospheric pressure. Such a high pressure could cause undesired mechanical deformations on the capacitor, therefore, a correction term must be included in (\ref{c}). A DCGT can measure temperature from 2.4\,K to 26\,K by using Helium ($^3$He, $^4$He), and the range can be further extended to above 100\,K with different gases (He, Ne and Ar) \cite{dcgt}. By far, the most accurate DCGT can measure temperature with an uncertainty of a few parts in $10^6$ \cite{te}. 
	
	The second electrical way of temperature measurement is using JNT \cite{jntr}. The JNT measures the temperature by averaging the statistical movement of electrons in an ohmic resistance, and $T$ is given by the mean square of the noise voltage ($V_{R}^2$) across a resistor $R$, i.e.
	\begin{equation}
	T=\frac{V_R^2}{4kR\Delta f},
	\end{equation}
	where $\Delta f$ is the noise voltage bandwidth. The JNT was used to measure $k$ at the triple point of water (0\,$^\circ$C or 273.15\,K). In this range, the effective noise voltage is on the order of $\mu$V. To keep the measurement signal stable and less affected by electronic noise, an in situ comparison with a quantum voltage noise reference generated by Josephson junctions is required. For low uncertainty measurements, the cross-talk of two channels needs to be evaluated and a long-term measurement is necessary. At 0\,$^\circ$C, the JNT can realize the kelvin with an accuracy of a few parts in $10^6$ \cite{jnt,jnt2}.
	
	%Joint efforts have been made between the National Institute of Metrology (NIM, China) and the National Institute of Standards and Technology (NIST, USA) to reduce the JNT measurement uncertainty \cite{jnt,jnt2}, and the best measurement result has an uncertainty of $2.7\times10^{-6}$ \cite{jnt,jnt2}.
	
	One of the major advances of the new SI system is that the base unit realization is no longer required to be fixed at a certain point. Since the JNT has a wide temperature measurement range (from below 1\,mK to above 1000\,K), the temperature measurement or realization is not necessarily set at the triple point of water. In fact, JNT has a greater advantage in measuring very low (e.g. $<5$\,K) and very high (e.g. $>800$\,K) temperatures. For low-temperature measurement, although the noise voltage signal is tiny, the measurement sensitivity can be compensated by using superconducting quantum interference devices (SQUIDs) \cite{squid}. SQUID-based JNT allows an accuracy below 1\% for mK temperature measurement \cite{jntr}. For high temperatures, the JNT measurement signal is much higher and will be no longer limited by electronic interferences.

	\section{Summary}
	The new SI revised the definition of the ampere from a deduction of the magnetic force with known geometry to fixing the numerical value of the elementary charge $e$. This revision eliminates the 1990 conventional electrical system and brings electrical measurements back to the SI system. In this survey, we discussed the most important impacts for electrical measurement activities in five different aspects: 1) the electrical unit scale adjustment, 2) the uncertainty variation for $\mu_0$ and $\varepsilon_0$, 3) the change of traceability and realization chain for electrical units, 4) the application of electrical measurements in mass and force calibration, and 5) the electrical applications for temperature measurement.
	
	Although there are minor adjustments for calibrations of some electrical quantities, in general, the implementation of the new SI revision leads to significant benefit and convenience to the electrical measurement and instrumentation society. All calibration service is now under SI and has long-term stability. Developments of quantum-based standards, sensors and components are greatly inspired. They simplify the traceability chain and lower the uncertainty for electrical measurements. It also shows the possibility of deeper application beyond the electricity area. For example, the mass (force), ranging from kilogram to microgram, will in the future be calibrated electrically by a Kibble balance or an electrostatic balance. The temperature realization or sensing can also be converted into electrical measurements, such as DCGT and JNT. In summary, a new era for electrical measurement has begun and everyone can take advantage of this SI reform.

	\section*{Appendix}
	
	The uncertainty change for some related physical constants before (in 2018) and after (at present) the 2019 SI revision is compared in Tab. \ref{tab4}. Note this update uses the 2017 CODATA adjustments for $h$, $e$, $N_A$, $k$ \cite{codata17}, and the latest adjustment for $\alpha$ \cite{codata18}.
	\begin{table*}[h]
		\centering
		\caption{Uncertainty change for physical constants before and after the 2019 SI revision. }
		\begin{tabular}{cccc}
			\hline \hline
			Quantity &    Symbol &    2018 ($u_r\times10^9$) &    2020 ($u_r\times10^9$)\\
			\hline
			International prototype of kilogram        &    $m(\mathcal{K})$            &0    &    10\\
			Permeability of free space                &    $\mu_0$                     &0    &    0.15\\
			Permittivity of free space                &    $\varepsilon_0$             &0    &    0.15\\
			Triple point of water                    &    $T_\textsc{TPW}$             &0    &370\\
			Molar mass of carbon-12                    &    $M$($^{12}$C)                &0    &0.30\\
			Planck constant                            &    $h$                         &10    &    0\\
			Elementary charge                        &    $e$                         &5    &0\\
			Boltzmann constant                        &    $k$                         &370&    0\\
			Avogadro constant                        &    $N_A$                         &10 &    0\\
			Molar gas constant                        &    $R$                         &370&    0\\
			Faraday constant                        &    $F$                         &5    &0\\
			Stefan-Boltzmann constant                &    $\sigma$                     &1480    &0\\
			Electron mass                            &    $m_e$                         &10    &0.30\\
			Atomic mass unit                        &    $m_u$                         &10    &0.30\\
			Mass of carbon-12                        &    $m$($^{12}$C)                &10    &0.30\\
			Josephson constant                        &    $K_J$                         &5    &0\\
			von Klitzing constant                    &    $R_K$                         &0.15&    0\\
			Fine-structure constant                    &    $\alpha$                     &0.15&    0.15\\
			$E = mc^2$ energy equivalent            &    J$\leftrightarrow$kg        &0    &    0\\
			$E = hc/\lambda$ energy equivalent        &    J$\leftrightarrow$m$^{-1}$    &10&    0\\
			$E = h\nu$ energy equivalent            &    J$\leftrightarrow$Hz        &10&    0\\
			$E = kT$ energy equivalent                &    J$\leftrightarrow$K            &370&    0\\
			1\,J=1\,(C/$e$)\,eV energy equivalent    &    J$\leftrightarrow$eV        &5&    0\\
			\hline \hline
		\end{tabular}
		\label{tab4}
	\end{table*}
	
	\section*{Acknowledgment}
	
	The authors would like to thank all reviewers for their valuable feedback to improve this survey. Shisong Li also wants to thank his colleagues from NIM, NIST and BIPM for years of fun towards and during the SI revision.  This work is supported in part by the National Key R\&D Program of China (Grant No.2016YFF0200102) and the Transforming Systems through Partnership (Grant No. TSPC1051) from the Royal Academy of Engineering, UK.

\end{document}